\documentstyle[11pt,newpasp,twoside,epsf]{article}
\def\edcomment#1{\iffalse\marginpar{\raggedright\sl#1\/}\else\relax\fi}
\reversemarginpar

\def\lesssim{\mathrel{\hbox{\rlap{\hbox{\lower4pt\hbox{$\sim$}}}\hbox{$<$}}}}
\def\gtrsim{\mathrel{\hbox{\rlap{\hbox{\lower4pt\hbox{$\sim$}}}\hbox{$>$}}}}

\begin{document}
\title{High Spatial Resolution Optical and Radio Imagery of the Circumbinary Environment}
 \author{Michael Bode}
\affil{Astrophysics Research Institute, Liverpool John Moores University, Twelve Quays House, Egerton Wharf, Birkenhead, CH41 1LD, United Kingdom}

\begin{abstract}
In this review, I concentrate on describing observations of spatially resolved emission in symbiotic stars at sub-arcsecond scales. In some of the closer objects, the highest resolutions discussed here correspond to linear dimensions similar to the supposed binary separation. A total of 17 stars well accepted as symbiotics are now observed to show sub-arcsecond structure, almost twice the number at the time of the last review in 1987. Furthermore, we now have access to HST imagery to add to radio  interferometry. From such observations we can derive fundamental parameters of the central systems, investigate the variation of physical parameters across the resolved nebulae and probe the physical mechanisms of mass loss and interactions between ejecta and the circumstellar medium. Suggestions for future work are made and the potential of new facilities in both the radio and optical domains is described. This review complements that by Corradi (this volume) which mainly considers the larger scale emission from the ionized nebulae of these objects.
\end{abstract}

\section{Introduction}

It is now fifteen years since Taylor's review of high resolution radio imagery of symbiotic stars was presented at the Torun conference (Taylor 1988). In that time, we have not only seen the sensitivity and frequency range of radio interferometers such as the VLA and MERLIN increase, but also, and perhaps more importantly, we have witnessed the advent of the Hubble Space Telescope. The HST has allowed ``routine'' sub-arcsecond imaging in optical wavebands. Taking the capabilities of each of these facilities in turn:

{\em The Very Large Array.} The VLA can operate in 8 frequency bands for continuum work from 0.07 to $43\;$GHz. In its largest configuration (A-array) with a baseline of $36\;$km, it can achieve $0.04\;$arcsecond resolution. At $5\;$GHz, its rms sensitivity is quoted as 50$\;\mu$Jy/beam in 10 minutes.

{\em MERLIN.} Possibly less well known outside the international radio community are the capabilities of the MERLIN array, centered on Jodrell Bank. MERLIN comprises an array of radio telescopes distributed across the UK with a maximum baseline of $217\;$km. It operates in three main bands at 1.7, 5 and $22\;$GHz. At the highest of these frequencies, MERLIN has a resolution of 0.008 arcseconds. It can achieve an rms sensitivity of 50$\;\mu$Jy/beam in a 12 hour track.

{\em HST.} The primary instruments that have provided the results discussed in this review are the Wide Field and Planetary Camera (WFPC2) and the Space Telescope Imaging Spectrograph (STIS). Both instruments can perform imaging through a range of filters. STIS can of course also perform long slit spectroscopy at a range of spectral resolutions. The operating wavelength range of these instruments is approximately 1200 to $11,000\;$\AA$\:$ and the spatial resolution of order $0.1\;$arcseconds. 

From this, it can be seen that for a ``typical'' symbiotic at a distance of 1kpc, MERLIN for example can theoretically resolve features with spatial extent $\sim 8\;$AU, which is only a few times larger than the binary separation in S-types (and of course, D-types have much larger separations). With observations down to these scales, in combination with modeling, and particularly if temporal variations can be monitored, we could potentially answer the following fundamental questions:

\begin{itemize}
\item{What are the orbital parameters (particularly of D-types)?}
\item{How is the nebular morphology linked to the parameters and activity of the central system, and can we test competing models of nebular formation?}
\item{Are we really seeing jets in some sources, and if so, what is their origin?}
\item{How is circumstellar dust distributed and can we disentangle the true value of interstellar extinction?}
\item{Are there additional systematic differences between different sub-types?}
\end{itemize}

It turns out that in some instances we can now make considerable headway on several of these points. However, as shown in the final section, the real breakthroughs will come with the next generation of ground-based facilities now in advanced planning or the early phases of construction.

\section{Review of Individual Objects}

Taylor (1988) discussed the limitations of current radio interferometers for detecting spatially extended emission in symbiotic stars. He then listed 10 objects where structure had been resolved at sub-arcsecond size scales using radio interferometry. In Table 1 below are listed all the objects generally accepted to lie within the symbiotic classification where such structure has been detected and the results subsequently presented in published sources. The list was compiled up to May 2002 and objects are shown in the order they are discussed in the text. The number of sources has thus expanded to 17 and includes HST and Australia Telescope (AT) detections. These of course did not figure in the Taylor review. In the case of the recurrent nova RS Oph, often included in the category of symbiotic, we were able to secure European VLBI Network observations during its 1985 outburst which illustrate what the highest spatial resolution radio observations can reveal (Taylor et al. 1989). Similarly, the Cambridge Optical Aperture Synthesis Telescope (COAST, see {\em e.g.} Baldwin \& Haniff 2001) is pioneering the use of optical interferometry and is discussed further in the final section.

\begin{table}
\caption{Spatially Resolved Emission at $\lesssim 1\;$arcsec
\footnotemark[1]}
\begin{tabular}{cccc}\tableline 
Object     &Sub-type&Detected Using          &Principal References \\ \tableline
R Aqr      &  D     &VLA, MERLIN, HST        &1,2,3\\
RX Pup     &  D     &VLA, HST                &4,5,6\\
HM Sge     &  D     &VLA, MERLIN, HST        &7,8,9,10,11\\
V1016 Cyg  &  D     &VLA, MERLIN, HST        &12,13\\
H1-36      &  D     &VLA                     &14\\
He2-104    &  D     &HST                     &15\\
BI Cru     &  D     &AT                      &7\\
He2-106    &  D     &AT                      &7\\
RR Tel     &  D     &AT                      &7\\
HD149427   &  D$\arcmin$    &AT, HST                 &7,16\\
Z And      &  S     &VLA, MERLIN             &7\\
AG Peg     &  S     &VLA, MERLIN             &7,17\\
AG Dra     &  S     &MERLIN                  &18\\
CH Cyg     &  S     &VLA, MERLIN, HST, COAST &19,20,21,22,23\\
SS73 96    &  S     &VLA                     &14\\
Hen 3-1383 &  ?     &VLA                     &14\\
RS Oph     &  RN    &VLA, EVN                &24,25\\ \tableline \tableline
\end{tabular}
\end{table}
\footnotetext[1]{References: 1. Dougherty et al. 1995; 2. Hollis et al. 1997; 3. Hollis et al. 2001; 4. Hollis et al. 1989; 5. Corradi \& Schwarz 2000; 6. Paresce 1990; 7. Kenny 1995; 8. Solf 1984; 9. Eyres et al. 1995; 10. Richards et al. 1999; 11. Eyres et al. 2001; 12. Watson et al. 2000; 13. Brocksopp et al. 2002; 14. Taylor 1988; 15. Corradi et al. 2001a; 16. Parthasarathy et al. 2000; 17. Kenny et al. 1991; 18. Mikolajewska 2002; 19. Taylor et al. 1986; 20. Crocker et al. 2001; 21. Eyres et al. 2002; 22. Corradi et al., 2001; 23. Crocker et al. 2002; 24. Hjellming et al. 1986; 25. Taylor et al. 1989}
We now proceed to discuss individual objects of particular interest in more detail.

\subsection{D-types}
\subsubsection{R Aquarii}

The large scale nebula surrounding R Aqr has been the subject of a great deal of detailed work over several decades (see Gon\c{c}alves, this volume). High resolution radio and optical (HST) observations of the inner nebula have been performed by several groups of recent years. 

\begin{figure}[t]
\plotfiddle{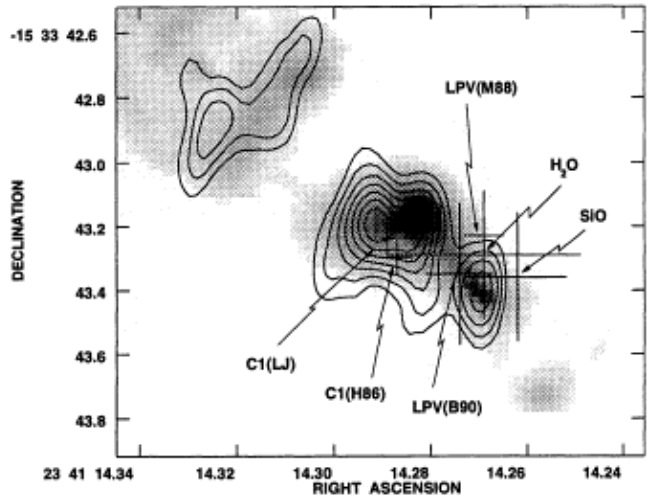}{2.5 in}{0}{75}{75}{-240 pt}{-190 pt}
\caption{$5\;$GHz MERLIN observations of R Aqr (Dougherty et al. 1995, contours) superimposed on HST $190\;$nm image of Paresce \& Hack (1994, grayscale). Positions of features identified by previous authors are indicated, including those of the Long Period Variable and maser sources. The central binary is located in the most westerly radio peak (``feature C1c'' - see Dougherty et al. for full details).}
\end{figure}

For example, Dougherty et al. (1995) observed R Aqr at 1.7 and $5\;$GHz with MERLIN yielding resolutions of 130 and $40\;$milliarcseconds (m.a.s.) respectively. In Fig. 1 are shown the $5\;$GHz results. Dougherty et al. argue that the symbiotic binary pair is located within what they designate as feature C1c. Spectral index mapping made possible by the quasi-simultaneous securing of observations at the two frequencies shows that the peak spectral index ($\alpha=1.6$) is located here. Assuming a spherically symmetric isothermal wind ($T=10^{4}\;$K) of constant outflow velocity ($v=30$ km s$^{-1}$) emitting via the free-free process ({\em i.e.} a Wright \& Barlow 1975 model) gives $\dot{M}=1.3\times10^{-8}\;$M$_{\odot}$ yr$^{-1}$ which is very low for a Mira variable. However, applying an ionization-bounded ($X_{STB}\sim0.04$) STB (Seaquist, Taylor \& Button, 1984) model with $v=10-30$ km s$^{-1}$ gives a much more reasonable $\dot{M}=5-15\times10^{-7}\;$M$_{\odot}$ yr$^{-1}$. 

Hollis, Pedelty \& Lyon (1997) have subsequently used the VLA at $43\;$GHz to map feature C1c at the frequency of the SiO maser (revealing a point source) and the adjacent continuum. The latter shows bipolar structure. If this is associated with emission pertaining to the binary components, and the separation of the peaks is the projected separation of the binary, then for $M\sim2.5-3$M$_{\odot}$ and $P\sim44$ years, $a\sim17-18$AU. Using preliminary orbital elements, Hollis et al. determine a distance to R Aqr of 195-206pc, very much in line with estimates from other methods. They also suggested that a monitoring campaign with the VLA at this frequency would obviously yield more precise system parameters. Most recently, Hollis et al. (2001) have used high spectral and spatial resolution observations with the VLBA in the $\nu=1, J=1-0$ SiO maser line. Their results are consistent with Keplerian (differential) rotation of the SiO maser shell.

\subsubsection{RX Puppis}

RX Pup is another symbiotic Mira system. Hollis et al. (1989) observed RX Pup at 5 and $15\;$GHz with the VLA in 1986. They found a possible northern elongation in the $5\;$GHz data and (more convincingly perhaps) found three components aligned  approximately E-W at the higher frequency on sub-arcsecond scales. The brightest of these is associated by these authors with the central star. They speculate that the other two features arise from a jet and that the morphology is reminiscent of that of R Aqr. Corradi \& Schwarz (2000) discuss optical spectroscopy of the region of an elongation at p.a. $15\deg$ extending to $\sim3.7\;$arcsec from the central star and first revealed in coronographic imaging by Paresce (1990). They confirm the presence of the feature, but find no evidence of this also being a jet. A tentative detection of elongated emission E-W was also made which may be related to the extended emission seen at $15\;$GHz by Hollis et al.

\subsubsection{HM Sagittae}

HM Sge is an outbursting symbiotic that underwent a major brightening in  1975. It has been observed in the radio by several workers at many epochs since then. VLA observations have shown extended structure on size scales from arcminutes to sub-arcsecond ({\em e.g.} Kenny 1995). As pointed out by Taylor (1988), the bipolar structure seen at the highest resolutions is very similar to that observed in V1016 Cyg. Indeed these two outbursting D-types share many similar characteristics (see below). Observations of HM Sge at $22\;$GHz by Li (1993) suggested ``rotation'' of the inner components (separation $\sim0.1\;$arcsec) implying a period of 65-118 years. 

In the optical, Solf (1984) performed heroic ground-based spatially-resolved spectroscopic observations. From these he deduced the existence of a bipolar mass flow in two lobes with separation $1.5\;$arcseconds (roughly E-W), two low velocity features lying N-S and separated by $0.2\;$arcseconds (deduced to be either a rotating ring or slowly expanding blobs) and a shell of emission around $0.5\;$arcseconds in diameter expanding at an intermediate velocity of $\simeq60\;$km s$^{-1}$. Solf then used these data to deduce a distance to HM Sge of approximately $400\;$pc.

\begin{figure}[t]
\plotfiddle{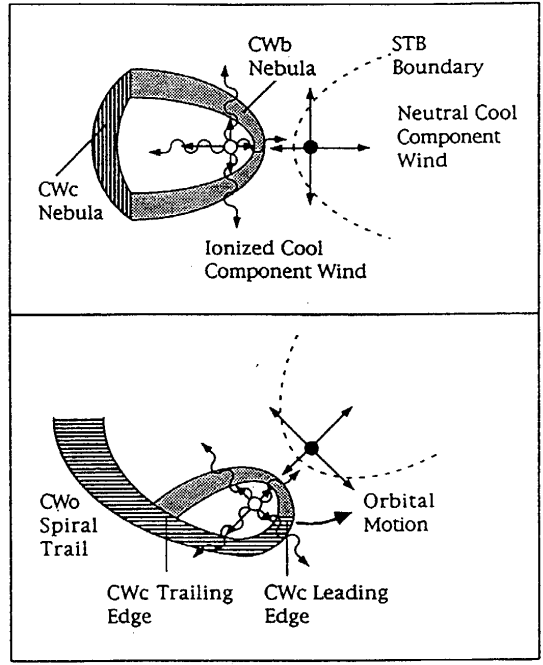}{3.8 in}{0}{40}{40}{-120 pt}{-15 pt}
\caption{Hybrid STB/CW model for HM Sge. Hot component denoted by $\circ$, cool component by $\bullet$. Orbital motion of the stars (bottom panel) leads to changing p.a. between the nebular features (from Kenny 1995 - see text for further details).}
\end{figure}

In his thesis, Kenny (1995) provides radio observations of several symbiotics, including HM Sge. He also develops models of their circumstellar environments, building on work of {\em e.g.} Kwok, Purton \& Fitzgerald (1978, concentric colliding winds - CWc), Girard \& Willson (1987, binary colliding winds - CWb) and Seaquist et al. (1984, STB). Figure 2 shows a composite colliding wind model coupled with an STB model, plus rotation. In the radio, such a confiiguration will naturally lead to peaks in the observed thermal emission which will rotate with the binary period. Projection effects due to orbital inclination  may then lead to changes in the apparent angular separation of the features.

Monitoring of HM Sge (and other symbiotics) has continued using both the VLA and MERLIN (Eyres et al. 1995; Richards et al. 1999), and most recently in combination with HST WFPC2 imaging (Eyres et al 2001). In Figure 3, Northern ({\em N}) and Southern ({\em S}) components of radio emission separated at $5\;$GHz by $\sim 150\;$m.a.s. are shown. Richards et al. again found an apparent rotation of radio components which is in line with the Kenny models and suggests a binary period $P = 90 \pm 20\;$years. Furthermore, combination of MERLIN 1.7 and $5\;$GHz observations allowed spectral index maps to be derived. Figure 4 clearly shows that although the emission near the supposed stellar position is dominated by optically thick thermal bremsstrahlung, as one moves E or W of the central source, the spectral index declines until non-thermal (synchrotron) emission is obviously dominant.

\begin{figure}[t]
\plotfiddle{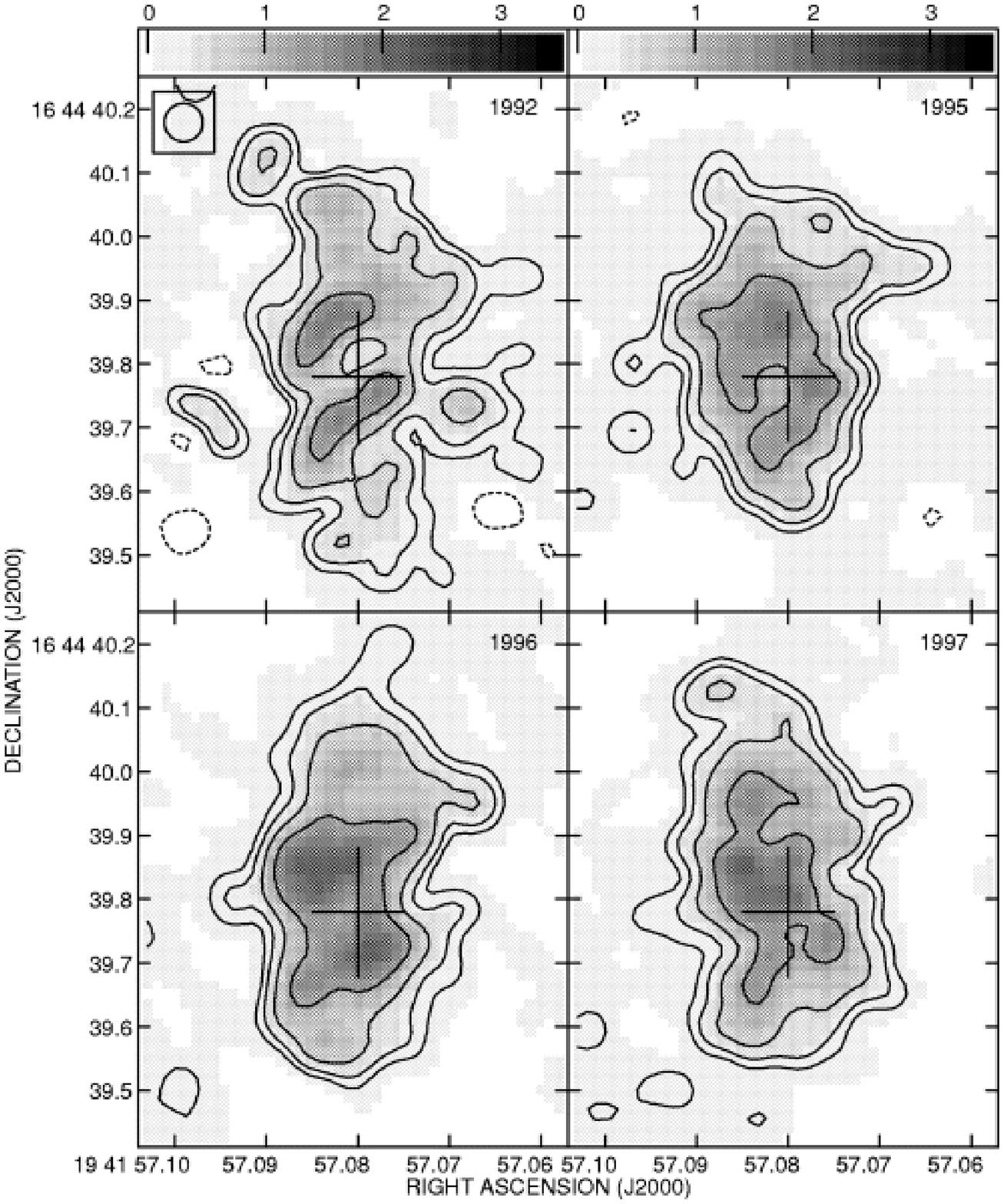}{4 in}{0}{50}{50}{-150 pt}{-50 pt}
\caption{MERLIN $5\;$GHz observations of HM Sge at four epochs (Richards et al. 1999). The mean optical position is also marked.}
\end{figure}

Aside from supernova remnants, spatially resolved, non-thermal emission is very rare in stellar sources and of course is the signature of particle acceleration and magnetic field enhancement in the circumstellar environment. Such effects would be expected in wind-ejecta interaction regions, and the anticipated presence of enhanced magnetic fields in the environs of the cool component is discussed in the review of Soker (this volume). From a simple model, it is  concluded that this emission should decline within decades, a prediction  which may help to explain the non-detection of such emission now in V1016 Cyg (see below). 

\begin{figure}[t]
\plotfiddle{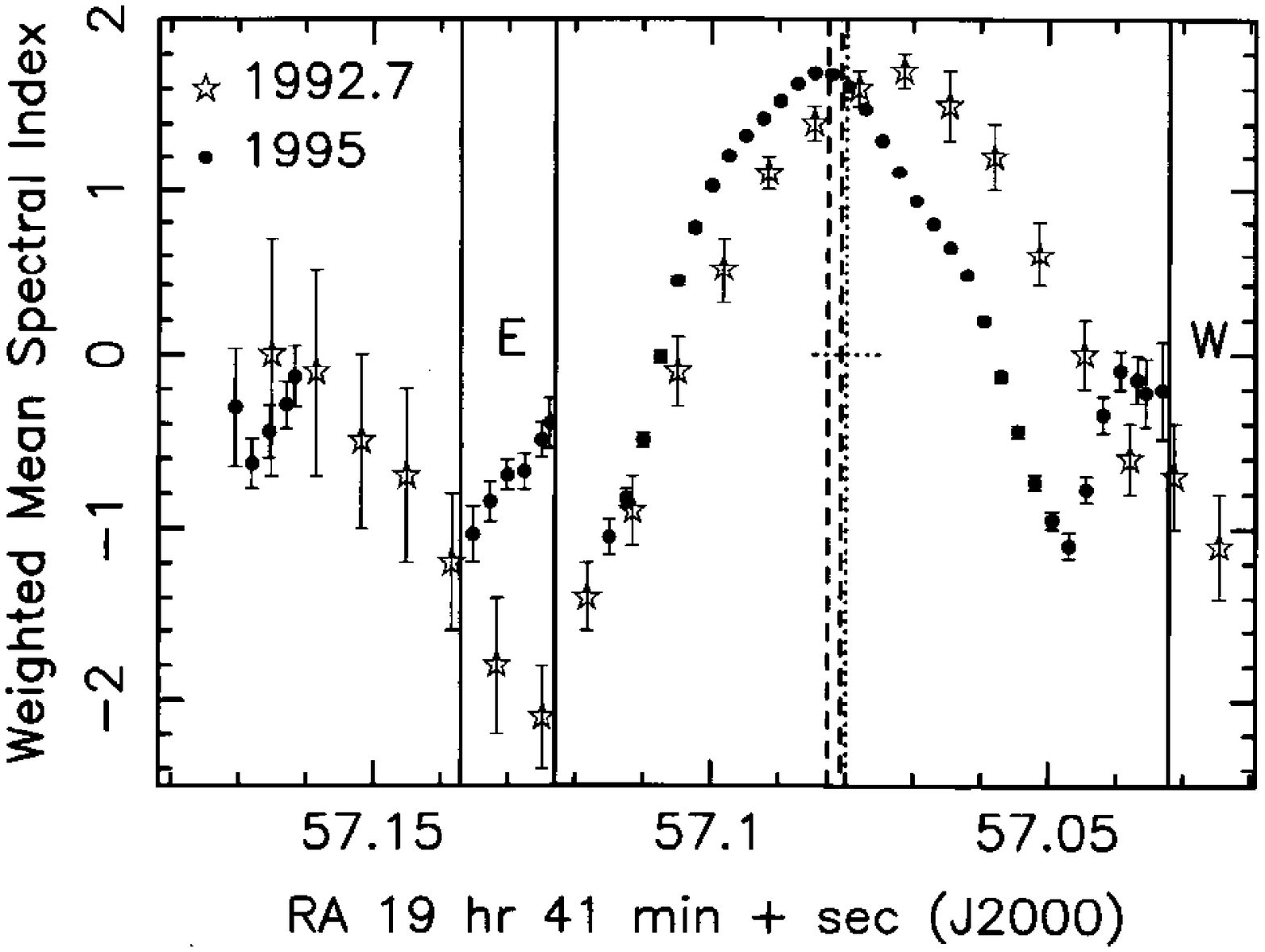}{3 in}{0}{50}{50}{-140 pt}{-90 pt}
\caption{HM Sge spectral index between 5 and $1.6\;$GHz at two epochs on an E-W slice. Vertical lines indicate approximate boundaries of the E and W components of non-thermal emission first detected by Eyres et al. (1995). Dotted and dashed lines indicate positions of optical and radio central features respectively (from Richards et al. 1999).}
\end{figure}

In HST GO program 8330, Eyres et al. (2001) used selected WFPC2 narrow band and broad band filters to image the nebular emission (a) at the wavelengths of diagnostic lines and (b) to locate the hot and cool continuum sources respectively. These observations were conducted on 1999 October 22. HM Sge had also been observed with the VLA in A-array at 8.56 and $23\;$GHz as part of the same study on 1999 September 26. Results are shown in Fig. 5. As well as the {\em N} and {\em S} features, a central peak ({\em C}) is indicated. This is where the central binary almost certainly lies. Several other larger scale nebular features are indicated. The [OIII] line ratio map clearly shows a ``wedge'' of cooler material to the SW, extending $\sim 0.5\;$arcseconds from the central regions of the nebula. Although it is dangerous to deduce precise temperatures from such ratios, the qualitative effects of temperature variation are certainly indicated. 

\begin{figure}[t]
\plotfiddle{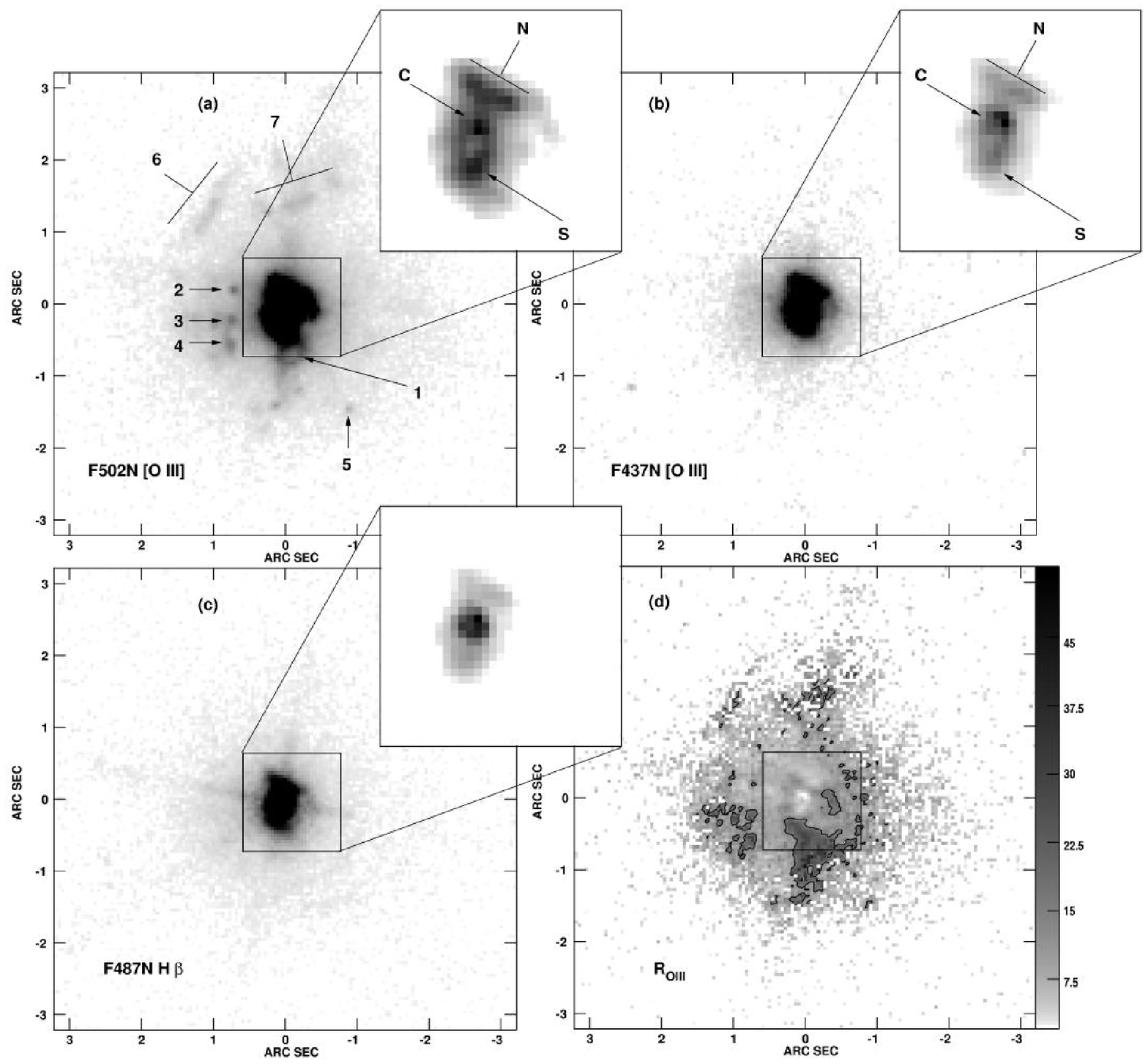}{3.75 in}{0}{55}{55}{-150 pt}{-10 pt}
\caption{HST images of HM Sge in three WFPC2 filters (a-c) with main resolved features indicated. (d) shows the ratio map of emission in (a) and (b) (Eyres et al. 2001).}
\end{figure}

A combination of H$\beta$ and radio continuum images have allowed the production of an extinction map of the innermost regions of the nebula. The principal results are that the peak of extinction lies in the direction of the cool wedge to the South of the central emission with a value $E_{B-V}\sim1$. The minimum value of extinction across the map is found to be $E_{B-V}\sim0.35$. If this truly is the minimum value, then it is a measure of the interstellar extinction to HM Sge and by comparison with large scale interstellar extinction maps, this places the object at $>700$pc. (NB: The $\lambda2200$ feature is not a good indicator of {\em total} extinction to a dusty source as it is only strongly correlated with the interstellar, and not the circumstellar component of extinction - see {\em e.g.} Bode 1987).

The broad band filters were chosen specifically to include only spectral regions expected to be dominated by emission from the hot and cool continua associated with the stellar components themselves. Using 2D gaussian fitting to dithered images, centroids of emission were calculated. These showed a positional offset of $40\pm9\;$m.a.s. at a p.a. of $130\deg\pm10\deg$ with the cool component being the more southerly. It should be noted that polarimetic work of Schmid et al. (2000, see also the more general review in this volume) gave a consistent p.a. of $123\deg$ for the line joining the stellar components. Taking $a = 25\;$AU (consistent with $P = 90$ years; $M = 2\;$M$_{\odot}$), and the fact that from the radio observations we expect the binary to have been observed at around the time of greatest elongation, gives $d = 625\pm140\;$pc. This is on the low side of previous estimates of the distance to HM Sge. Accurate positions of emission peaks were also derived through the narrow band (line) filters. For example, this placed the peak of the HeII emission between the stellar components, consistent with the model of Nussbaumer \& Vogel (1990). Results consistent with expectations are also found for V1016 Cyg and CH Cyg (see below). However, because of the potential importance of these results, investigations are still continuing to determine whether the offsets in positions could be due to a subtle systematic effect. Unfortunately, at the present time it is proving difficult to arrive at a definitive answer (see also Brocksopp et al., this volume).

\subsubsection{V1016 Cyg}

As noted above, this object is the virtual twin of HM Sge in many ways. V1016 Cyg also underwent a nova-like outburst (in 1965), it shows a bipolar radio morphology on sub-arcsecond scales (see {\em e.g.} Taylor 1988) and Solf (1983) had deduced a very similar optical morphology to HM Sge. Recent MERLIN $5\;$GHz observations over a period of several years have indicated that as with HM Sge, there is an apparent rotation of the central features (Watson et al.  2000  - see also Fig. 6). Comparison of Figs. 3 and 6 emphasizes how remarkably similar the radio morphologies of these two objects are at present. Unlike HM Sge however, spectral index mapping did not reveal any extended non-thermal component to the emission.

\begin{figure}[t]
\plotfiddle{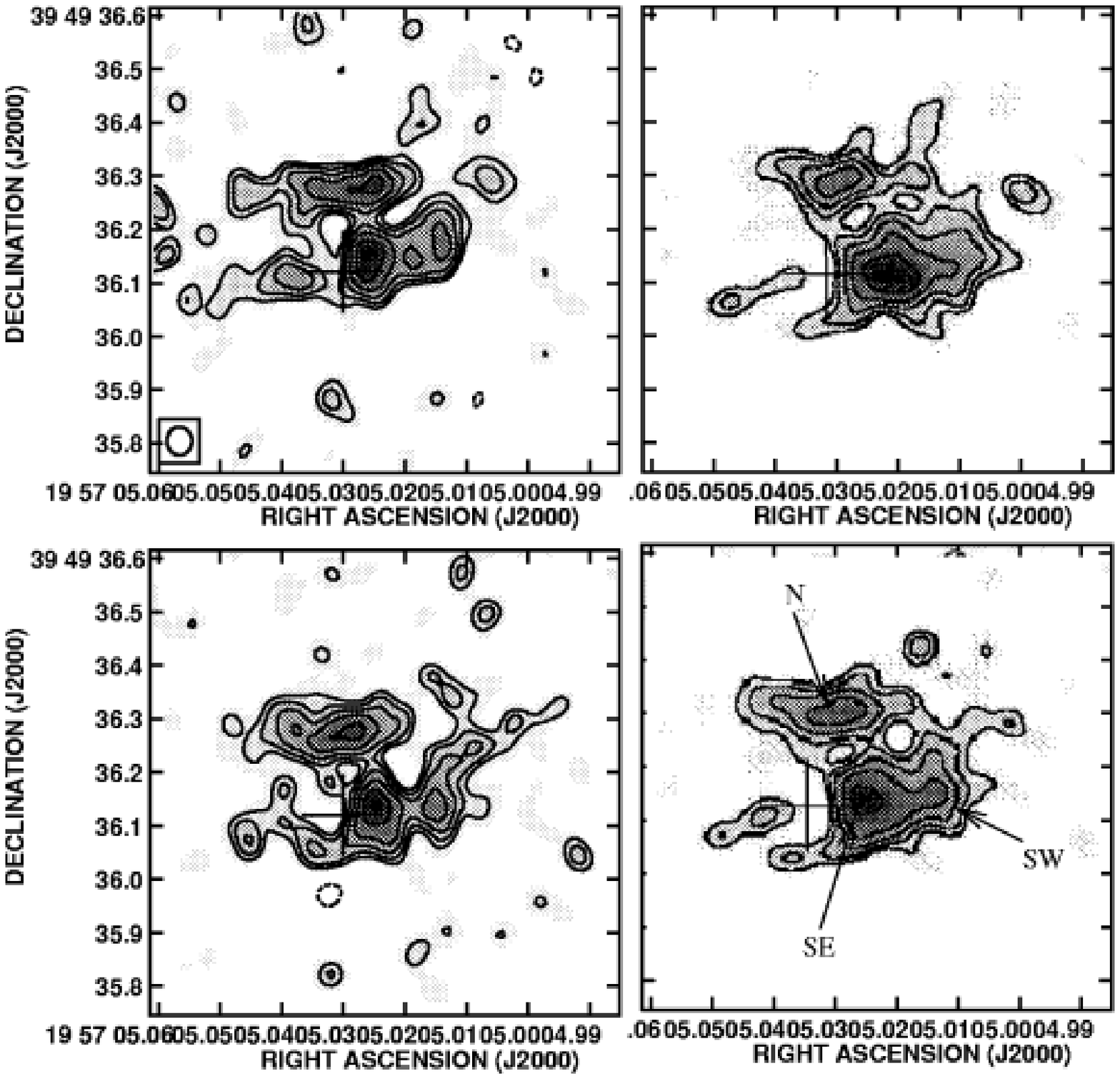}{3.75 in}{0}{60}{60}{-200 pt}{-95 pt}
\caption{MERLIN maps at 5 GHz of V1016 Cyg from 1992 July 21 (top left), 1995 June 25 (bottom left) and 1996 November 9/1997 February 3 (top right). Also shown, with main resolved features indicated, is a composite image of all four epochs (Watson et al. 2000).}
\end{figure}

V1016 Cyg was also a target of HST GO program 8330. Again, a range of filters was used to map emission at various diagnostic line wavelengths and to locate the positions of the binary components. VLA observations at $23\;$GHz in A-array were made within 6 weeks of those with HST. These data reveal other similarities with HM Sge in terms of apparent binary separation and position angle on the sky. In addition, the sub-arcsecond nebula was observed with STIS as part of a spectroscopic SNAP program (8188) with HST. The differences between the two objects ({\em e.g.} the greater observable structure in HM Sge; the presence or otherwise of extended non-thermal emission) can be explained in terms of the probability that V1016 Cyg is more distant, and certainly had an earlier outburst.  A full discussion of the results of all these recent radio and HST observations is given in Brocksopp et al. (2002 - see also contribution in this volume).

\subsection{S-types}

\begin{figure}[t]
\plotfiddle{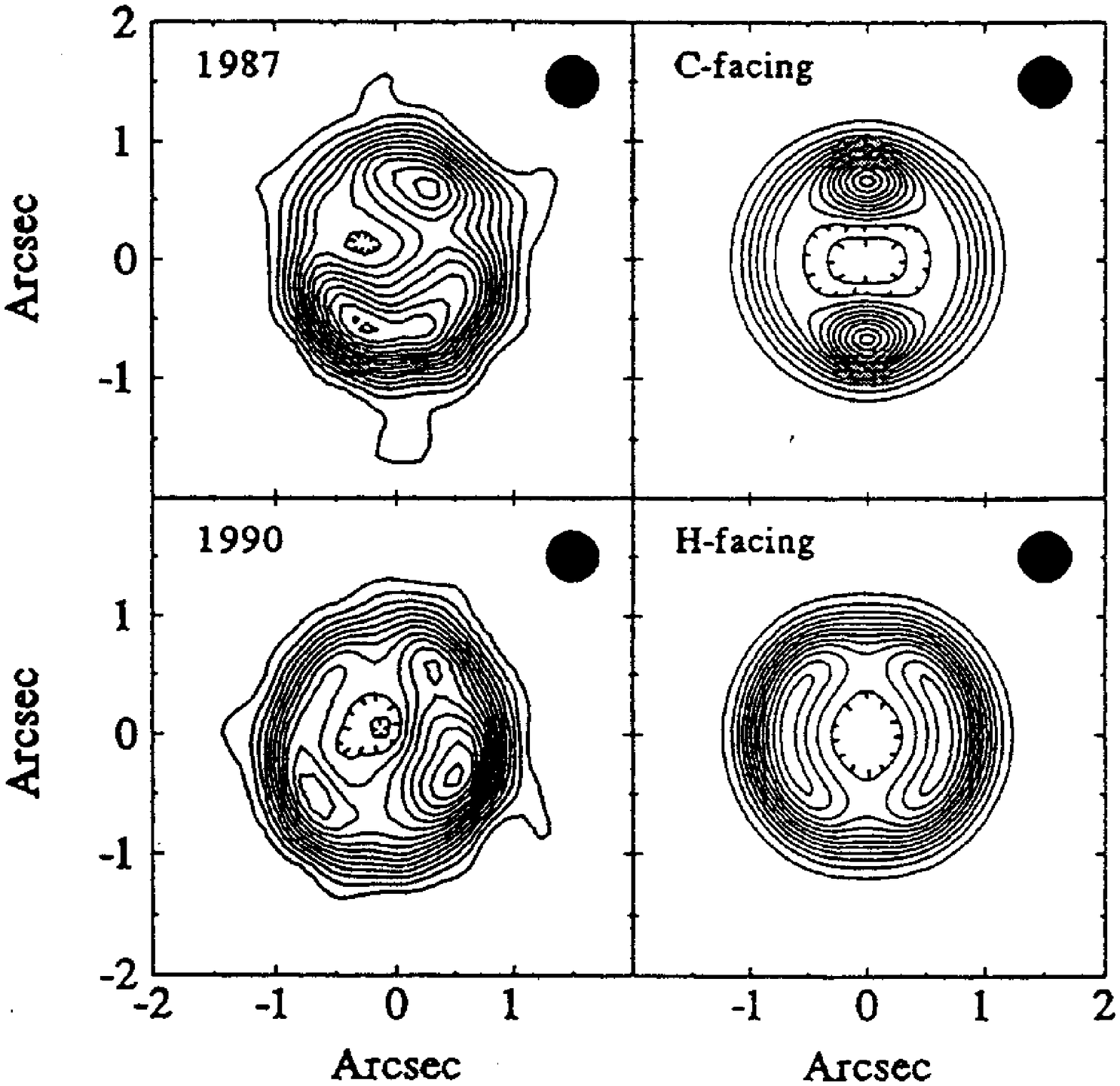}{4 in}{0}{50}{50}{-170 pt}{-50 pt}
\caption{VLA $5\;$GHz images of AG Peg (left) compared with model images (from Kenny 1995 - see text for further details).}
\end{figure}

\subsubsection{Z Andromedae}

Z And is an active system that has undergone several recorded outbursts (see {\em e.g.} Fernandez-Castro et al. 1995). Kenny (1995) discusses near-simultaneous MERLIN and VLA observations in 1991-1992. These reveal a central source extended by $\sim0.1\;$arcsec E-W. Kenny fitted these observations with a simple STB model plus a ``no recombination'' radius of $70\;$AU (this being the radius from the central system at which ionized gas does not have time to recombine before the ionization cone around the hot component sweeps through again as the binary rotates). The best fit model (both to the images and radio light curves) gave $\dot{M}/v = 9.5\times10^{-9}\;$M$_{\odot}$ yr$^{-1}$(km s$^{-1}$)$^{-1}$, $L_{ph} = 2.6\times10^{45}\;$s$^{-1}$, $X_{STB} = 0.36$ and $\theta_{a} = 30\deg$ for $a = 2\;$AU and $d = 1.1\;$kpc (where $L_{ph}$ is the hot component ionizing photon flux, $v$ is the red giant wind velocity and $\theta_{a}$ is the opening angle of the cone of ionized material). In addition to the central extended source, a ``blob'' of emission $\sim0.1\;$arcsec to the North was suggested as possibly being collimated ejecta from the 1984/85 outburst of the system.

\subsubsection{AG Pegasi}

AG Peg is an archetypal symbiotic nova that underwent a major outburst in  ca.1850. It shows structure in the radio from arcminute to sub-arcsecond scales (Kenny, Taylor \& Seaquist 1991). Kenny (1995) reports MERLIN plus VLA observations. He has modeled the emission via colliding winds with $v_{hot} = 900\;$km s$^{-1}$, $v_{cool} = 10\;$km s$^{-1}$, $\dot{M}_{cool} = 2.1\times10^{-7}\;$M$_{\odot}$yr$^{-1}$, $i = 65\deg$ and projected polar axis p.a. $= -15\deg$ for $a=2.5\;$AU, $d=600\;$pc (see Fig.7). Variation in the resolved images over the three year period is deduced to be due to orbital motion and a factor $\sim$3 decrease in mass loss rate from the hot component to 0.1$\dot{M}_{cool}$ in 1990.

\subsubsection{AG Draconis}

Mikolajewska (2002, also this volume) has drawn attention to the fact that the MERLIN images of AG Dra in Ogley et al. (2002) show two peaks of emission, the line joining which is approximately aligned with the binary axis deduced from spectropolarimetric observations of Schmid \& Schild (1997). Assuming an ejection velocity of the order of the escape velocity from the hot component, a system distance of $\sim2.5\;$kpc and an orbital inclination $\sim120\deg$ gives a time for ejection of this material (if in the form of jets or ``bullets'') consistent with the active phase of the object between 1995 and 1998.

\subsubsection{CH Cygni}

CH Cyg has also undergone several outbursts since first being designated as symbiotic in 1964 (see also contributions by Tomov and Leedjaerv, this volume). Radio observations in 1984 following the end of its largest outburst to-date revealed the presence of a rapidly expanding bipolar structure, conjectured to be a relatively high velocity jet (Taylor, Seaquist \& Mattei 1986; Taylor 1988).  

Crocker et al. (2001) report a study of new and archival VLA observations at several frequencies obtained between 1985 and 1999. The central region of the resolved source is always dominated by thermal emission. However, spectral index maps clearly show that emission from the extended regions is unambiguously non-thermal and that it does indeed have a jet-like morphology. Again, the presence of synchrotron emission implies particle acceleration and magnetic field enhancement in shocked regions, probably arising as the high velocity ejecta interact with a pre-existing circumstellar medium. Magnetic field strengths of order 3mG are derived in the jets.

\begin{figure}[t]
\plottwo{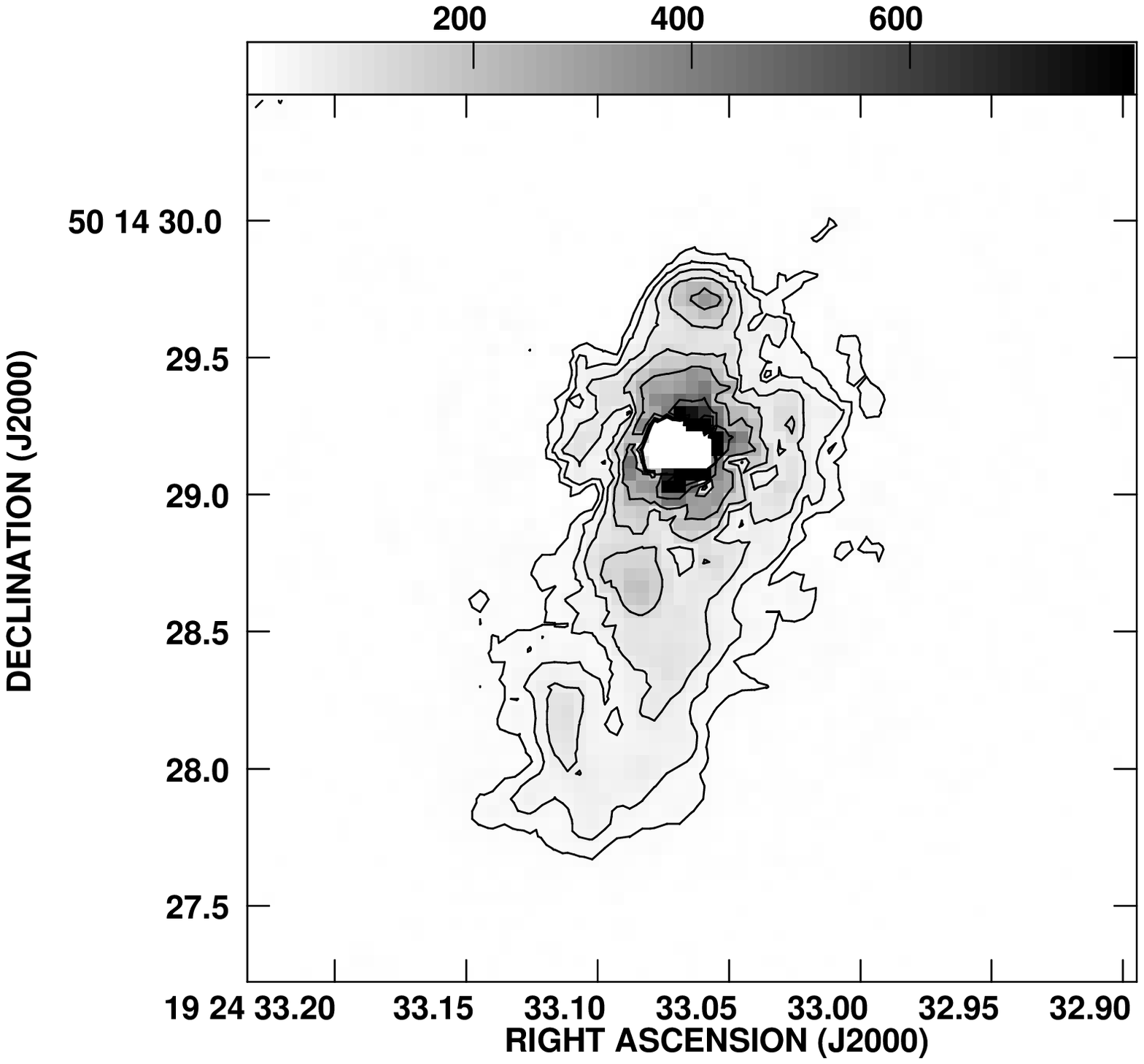}{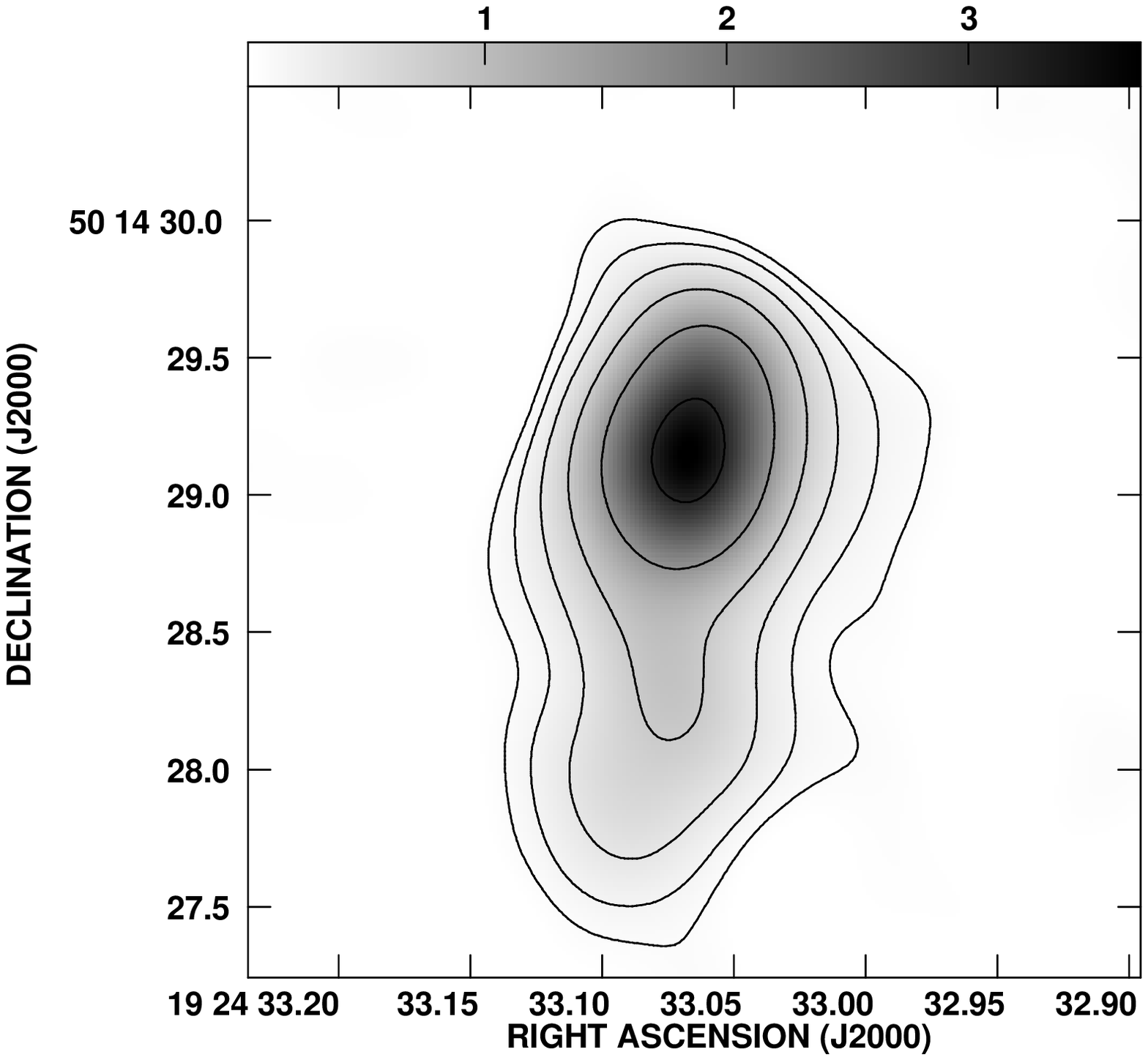}
\caption{HST WFPC2 [OIII] image of CH Cyg on 1999 August 12 (left) compared with VLA $5\;$GHz image obtained on 1999 September 26 (Eyres et al. 2002).}
\end{figure}

Contemporaneous HST and VLA imaging was secured by Eyres et al. (2002) in August and September 1999 (see Fig. 8). The HST observations were part of GO program 8330 alluded to above and followed the same format. Further HST images were obtained by Corradi et al. (2001b) in 1999 October and complement those of Eyres et al. The surface brightness contours of the contemporaneous HST and VLA images show broad agreement with respect to the overall morphology. The fact that the extended radio emission contained a significant non-thermal component meant however that results of extinction mapping in the nebula using the technique mentioned above are viewed with some added caution. At the time of the Eyres et al. observations, CH Cyg was predicted to have its hot component eclipsed by the cool giant on the proposed long-period ($14.5\;$ yr) outer orbit of the suspected triple system. The fact that there was no difference in derived positions for the hot and cool components utilizing the 2D gaussian fitting  methods described above was then consistent with eclipse (as were various spectroscopic and photometric changes observed from the ground - Eyres et al. 2002).

\begin{figure}
\plotfiddle{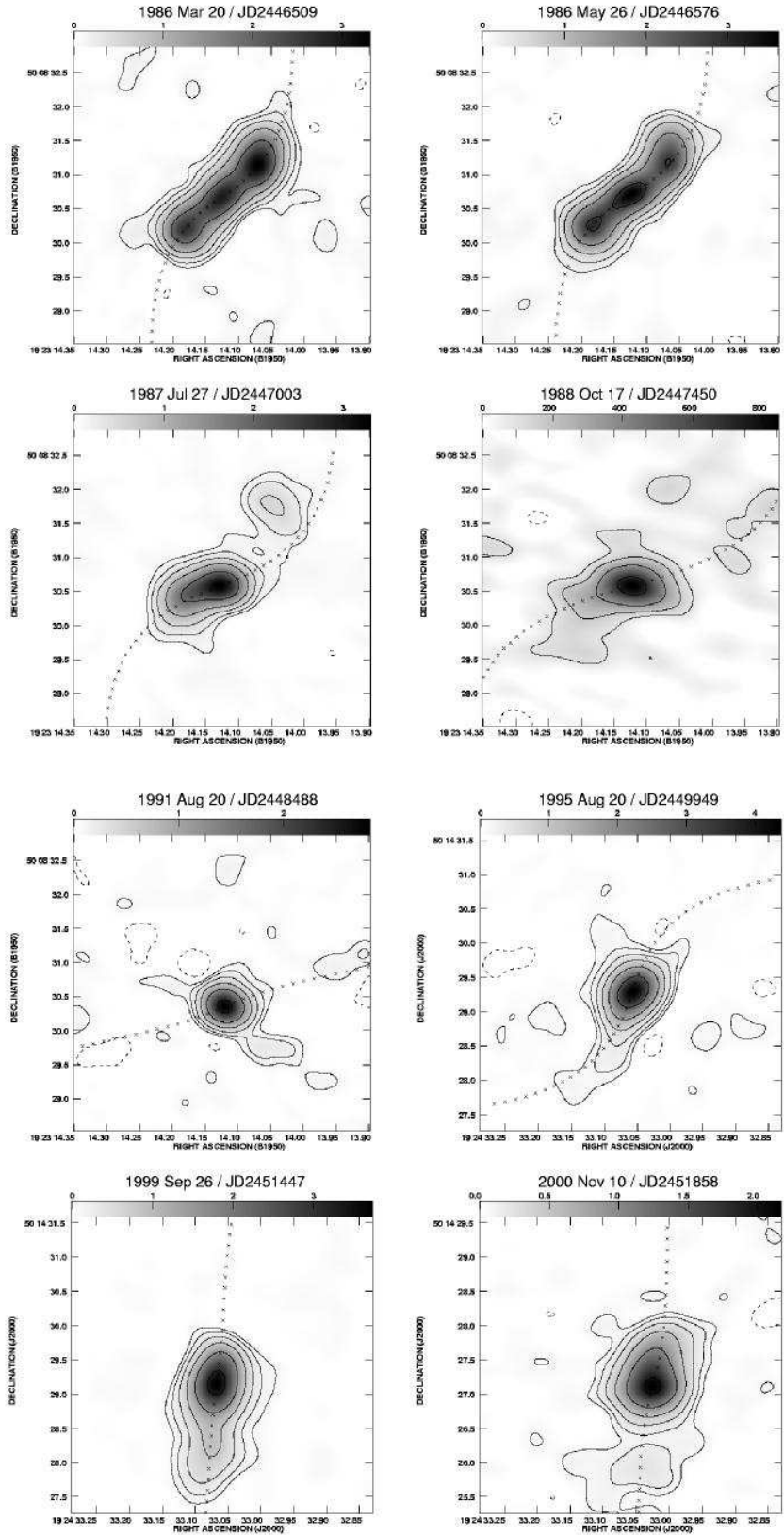}{8.5 in}{0}{80}{80}{-200 pt}{-22 pt}
\caption{VLA $5\;$GHz images of CH Cyg from 1986-2000 compared with a simple ballistic model (dotted line - see Crocker et al. 2002). Each box is $\sim4\;$arcsec on a side.}
\end{figure}

It is obvious from comparing the VLA observations reported in Taylor et al. (1986) and Eyres et al. (2002) that the position angle of ejection had changed markedly. This has been studied in some detail by Crocker et al. (2002). Using both radio and HST imagery, the evolution of the morphology of the inner regions was fitted using a precessing jet model adapted from that used for SS433 (for non-relativistic expansion velocities). The best fit model has precession period $P=6519\pm153\;$days; the position angle on the sky of the axis of precession $\theta_{mid}=140\deg\pm1\deg$; the cone opening angle  $\theta_{0}=35\deg\pm1\deg$; the velocity of material ejected in the jet $v_{0}=1263\pm18\;$km s$^{-1}$, and the inclination of the precession axis to the line of sight $i=88\deg\pm1\deg$. Figure 9 shows the results of the model fit for several epochs. Systematic optical spectroscopic monitoring with sufficient spectral resolution and conducted over a sufficiently large time scale would be very useful in exploring the veracity of this model.  

Crocker et al. (2002) note that there are epochs where anomalous knots appear in the emission which may be older, slower-moving ejecta, or possibly jet material that has been disrupted through transverse interaction with the surrounding medium. However, the general behavior is dominated by ejecta that seem to behave in a consistent manner with the above model. As well as providing evidence of an efficient collimation mechanism, the authors also consider causes of the precession of the jet. The most favored is warping of the collimating accretion disk due to the presence of a strong magnetic field emanating from the white dwarf. Other possible mechanisms yield precession periods that are several orders of magnitude longer than that observed. 

Most recently Young et al. (in preparation) have demonstrated the potential of optical interferometry in probing the innermost regions of symbiotics. They have used COAST to observe CH Cyg at $905\;$nm. At the time of the observations,  COAST comprised five $40\;$cm telescopes with a maximum $48\;$m baseline (recently increased to $67\;$m) and is situated in Cambridge, UK. The resolution of the array was $9\;$m.a.s., equivalent to $2.4\;$AU at the distance of CH Cyg. Between 1999 and 2001, they found persistent elongation of the central source with possible rotation with a period similar to that of the proposed outer giant's orbit. Puzzlingly, they did not appear to detect the emergence of the outer giant following its eclipse of the hot component in 1999. Although further work needs to be done to come to firm conclusions, these results demonstrate the impact that future facilities will have on our understanding of the fundamental parameters of symbiotic systems (see below).

\section{Concluding Remarks}

\subsection{Progress Report}

As I hope the above summary of results for individual objects demonstrates, a good deal of progress has been made since the time of the last review to answer some of the major questions posed in the introduction. However, much more work remains. In summary:

\begin{itemize}
\item{Observations and modeling of resolved emission at sub-arcsecond scales are consistent with much greater binary separations in D-types compared to S-types. It is suggested that changes in radio morphology may be directly linked to orbital periods, which may be particularly important for our understanding of D-type systems.}
\item{We have shown that the nebular morphology can be directly linked to parameters of the central system. However, there is no doubt that we are dealing with a complex and diverse set of phenomena in a relatively heterogeneous class of objects and that the models discussed here are essentially ``first order''. The observational data set now warrants the application of more detailed and physically realistic models such as those discussed by Nussbaumer and Kilpio (this volume).}
\item{In D-types, the presence of ejection via a jet is well established in R Aqr, but more work is needed to establish this beyond doubt in RX Pup. Similarly, for the S-types, it now appears that the expanding features resolved in the radio in CH Cyg are indeed jets whose axis of ejection may be undergoing precession. The non-thermal emission detected in this object (and in the D-type HM Sge) confirms, amongst other things, the presence of enhanced magnetic fields in extended regions. A central accretion disk may act as the jet collimator whilst the presence of a magnetic white dwarf may explain apparent precession of the jet ejection axis and its occasional one-sidedness, plus being a plausible source of the magnetic field inferred to be swept out in the jets. In Z And and AG Dra, there may be some evidence for jet ejection from direct imaging, but the evidence is as yet nowhere near as compelling as for CH Cyg. As always, we need to be wary of ascribing the word ``jet'' too early to miscellaneous ``blobs'' of emission.}
\item{The potential has been demonstrated for probing the distribution of circumstellar dust in a few objects using a combination of radio and narrow-band HST imagery. Disentangling interstellar and circumstellar contributions to the visual extinction is obviously of great importance in determination of the most fundamental of parameters, the distance.}
\item{As to the final question posed in the introduction, we need a much larger sample of objects to be resolved at sub-arcsecond scales to draw any compelling conclusions about further systematic differences between the sub-types, and hence possibly to derive important additional information on their evolution. However, the advent of new facilities promises that such large samples are less than a decade away.}
\end{itemize}

\subsection{For the Future}

In the near-term, we should continue our radio monitoring program of selected objects (in particular HM Sge, V1016 Cyg and CH Cyg). In the optical, spatially-resolved spectroscopy of circumstellar nebulae with STIS on HST shows great potential, but has not been fully exploited. Such observations would turn the semi-qualitative initial results from narrow band imaging into more secure, quantitative measures. As already noted, we should also begin to apply more sophisticated models to the full multi-frequency data set.

Over the next few years, and certainly before another fifteen have passed, a new generation of ground-based optical and radio facilities will be available. In the radio, projects to enhance the performance of both the VLA (EVLA) and MERLIN (e-MERLIN) are now underway. These instruments have the potential to revolutionize the subject. For example, based on the survey of Seaquist, Krogulec \& Taylor (1993), e-MERLIN, with up to $\sim50$ times the sensitivity of the present array, will be able to detect extended emission in $\gtrsim4$ times as many symbiotics as at present, with resolutions as high as $\sim10\;$ m.a.s. Around half of these will be S-types (see www.merlin.ac.uk/e-merlin/ for further details of the e-MERLIN project).

In the optical, great strides are being made in the application of adaptive optics on large telescopes. In addition, we have already seen that interferometric arrays such as COAST can produce significant results, but at present only on a very limited set of very bright objects such as CH Cyg. Arrays on better sites, with larger total collecting area and longer baselines have already been, or are in the process of being built (Baldwin \& Haniff 2001). However, more ambitious than these is the Large Optical Array (LOA) which is the next generation COAST. Importantly, the LOA would have a relatively large number of moderate aperture telescopes, enabling ``snapshot'' optical imaging of sources in an analogous fashion to the VLA in the radio. In its original conception the LOA comprises 15 $\times\;1.2\;$m optical telescopes distributed over an area of $500\;$m $\times$ $500\;$m and operating between 0.6 and 2.4$\;\mu$m in wavelength. This gives a resolution equivalent to $0.4\;$AU at $1\;$kpc and the ability to map sources to J,H,K $\sim 14$. Thus the LOA could probe to within the central binaries of S-type symbiotics and follow morphological changes around their binary orbits. Although the full LOA is unlikely to be built with purely UK funding, it is almost certain that something approaching its capabilities will be built through international collaboration within the next decade.

\acknowledgments
I would like to thank all my collaborators over the years (particularly my students and postdoc's who end up doing most of the hard work!); Dr John Young of MRAO, Cambridge, for provision of COAST results on CH Cyg; Dr Andy Newsam and Mr David Hyder of the Astrophysics Research Institute for assistance with the figures, and finally the conference organizers for facilitating my attendance at such a splendid meeting.


\begin{references}
\reference{Baldwin, J.E., \& Haniff, C.A. 2001, Phil. Trans. A., 360, 969}
\reference{Bode, M.F. 1987, in Dust in the Universe, ed. M.E. Bailey \& D.A. Williams (Cambridge: CUP), 73}
\reference{Brocksopp, C., Bode, M.F., Eyres, S.P.S., Crocker, M.M., Davis, R.J., \& Taylor, A.R. 2002, \apj, 571, 947}
\reference{Corradi, R.L.M., \& Schwarz, H.E. 2000, \aap, 363, 671}
\reference{Corradi, R.L.M., Livio, M., Balick, B., Munari, U., \& Schwarz, H.E. 2001a, \apj, 553, 211}
\reference{Corradi, R.L.M., Munari, U., Livio, M., Mampaso, A., Gon\c{c}alves, D.R., \& Schwarz, H.E. 2001b, \apj, 560, 912}
\reference{Crocker, M.M., Davis, R.J., Eyres, S.P.S., Bode, M.F., Taylor, A.R., Skopal, A., \& Kenny, H.T. 2001,  \mnras, 326, 781}
\reference{Crocker, M.M., Davis, R.J., Spencer, R.E., Eyres, S.P.S., Bode, M.F., \& Skopal, A. 2002, \mnras, in press}
\reference{Dougherty, S.M., Bode, M.F., Lloyd, H.M., Davis, R.J., \& Eyres, S.P.S. 1995, \mnras, 272, 843}
\reference{Eyres, S.P.S., Kenny, H.T., Cohen, R.J., Lloyd, H.M., Dougherty, S.M., Davis, R.J., \& Bode, M.F. 1995, \mnras, 274, 317}
\reference{Eyres, S.P.S., Bode, M.F., Taylor, A.R., Crocker, M.M., \& Davis, R.J. 2001, \apj, 551, 512}
\reference{Eyres, S.P.S., Bode, M.F., Skopal, A., Crocker, M.M., Davis, R.J., Taylor, A.R., Teodorani, M., Errico, L., Vittone, A.A., \& Elkin, V.G. 2002, \mnras, in press}
\reference{Fernandez-Castro, T., Gonzalez-Riestra, R., Cassatella, A., Taylor, A.R., \& Seaquist, E.R. 1995, \apj, 442, 366} 
\reference{Girard, T., \& Willson, L.A. 1987, \aap, 183, 247}
\reference{Hjellming, R.M., van Gorkum, J.H., Taylor, A.R., Seaquist, E.R., Padin, S., Davis, R.J., \& Bode, M.F. 1986, ApJL, 305, L71}
\reference{Hollis, J.M., Yusuf-Zadeh, F., Cornwell, T.J., Oliversen, R.J., Michelitsianos, A.G., \& Kafatos, M. 1989, \apj, 337, 514}
\reference{Hollis, J.M., Pedelty, J.A., \& Lyon, R.G., 1997, \apj, 482, L85}
\reference{Hollis, J.M., Boboltz, D.A., Pedelty, J.A., White, S.M., \& Forster, J.R., 2001, \apj, 559, L37}
\reference{Kenny, H.T., Taylor, A.R., \& Seaquist, E.R. 1991, \apj, 366, 549}
\reference{Kenny, H.T. 1995, PhD Thesis, University of Calgary}
\reference{Kwok, S., Purton, C.R., \& Fitzgerald, M.P. 1978, \apj, 219, L125}
\reference{Li, P.S. 1993, MSc Thesis, University of Calgary}
\reference{Mikolajewska, J. 2002, \mnras, in press}
\reference{Nussbaumer, H., \& Vogel, M. 1990, \aap, 236, 117}
\reference{Ogley, R.N., Chaty, S., Crocker, M.M., Eyres, S.P.S., Kenworthy, M.A., Richards, A.M.S., Rodriquez, L.F., \& Stirling, A.M. 2001, \mnras, 330, 772}
\reference{Paresce, F. 1990, \apj, 357, 231}
\reference{Paresce, F., \& Hack, W. 1994, \aap, 287, 154}

\reference{Parthasarathy, M., Garc\'{i}a-Lario, P., Pottasch, S.R., de Martino, D., \&  Surendiranath, R. 2000, \aap, 355, 720} 
\reference{Richards, A.M.S., Bode, M.F., Eyres, S.P.S., Kenny, H.T., Davis, R.J., \& Watson, S.K. 1999, \mnras, 305, 380}
\reference{Seaquist, E.R., Taylor, A.R., \& Button, S. 1984, \apj, 284, 202}
\reference{Seaquist, E.R., Krogulec, M., \& Taylor, A.R. 1993, \apj, 410, 260}
\reference{Schmid, H.M., \& Schild, H. 1997, \aap, 321, 791}
\reference{Schmid, H.M., Corradi, R., Krautter, J., \& Schild, H. 2000, \aap, 355, 261}
\reference{Solf, J. 1983, ApJL, 266, L113}
\reference{Solf, J. 1984, \aap, 139, 296}
\reference{Taylor, A.R., 1988 in The Symbiotic Phenomenon, ed. J. Mikolajewska et al. (Dordrecht: Kluwer), 77}
\reference{Taylor, A.R., Seaquist, E.R., \& Mattei, J. 1986, Nature, 319, 38}
\reference{Taylor, A.R., Davis, R.J., Porcas, R., \& Bode, M.F. 1989, \mnras, 237, 81}
\reference{Watson, S.K., Eyres, S.P.S., Davis, R.J., Bode, M.F., Richards, A.M.S., \& Kenny, H.T.,  2000, \mnras, 311, 449}
\reference{Wright, A.E., \& Barlow, M.J. 1975, \mnras, 170, 41}
\end{references}
\end{document}